\documentclass[apsrev4-1,prl,superscriptaddress,longbibliography,twocolumn]{revtex4-1}
\usepackage{colortbl}
\usepackage{color}
\usepackage{amsfonts,amsmath,amssymb}
\usepackage[dvipsnames]{xcolor}
\usepackage{tabularx,hhline,tikz,multirow,array}
\usepackage{bm,enumitem}
\usepackage{dcolumn,stmaryrd}
\usepackage[colorlinks=true,pdfborder={0 0 0},linkcolor=red,citecolor = blue]{hyperref}

\newcommand{\Dp}{\mathbf{D}}
\newcommand{\Sp}{\mathbf{S}}

\allowdisplaybreaks

\begin{document}

\title{Skyrmion fluid and bimeron glass protected by a chiral spin liquid on a kagome lattice}

\author{H. Diego Rosales}
\affiliation{Instituto de F\'isica de L\'iquidos y Sistemas Biol\'ogicos, CONICET, Facultad de Ciencias Exactas, Universidad Nacional de La Plata, 1900 La Plata, Argentina}
\affiliation{Departamento de F\'isica, FCE, UNLP, La Plata, Argentina}
\affiliation{Departamento de Ciencias B\'asicas, Facultad de Ingenier\'ia, UNLP, La Plata, Argentina}

\author{Flavia A. G\'omez Albarrac\'in}
\affiliation{Instituto de F\'isica de L\'iquidos y Sistemas Biol\'ogicos, CONICET, Facultad de Ciencias Exactas, Universidad Nacional de La Plata, 1900 La Plata, Argentina}
\affiliation{Departamento de F\'isica, FCE, UNLP, La Plata, Argentina}
\affiliation{Departamento de Ciencias B\'asicas, Facultad de Ingenier\'ia, UNLP, La Plata, Argentina}

\author{Pierre Pujol}
\affiliation{Laboratoire de Physique Th\'eorique, CNRS and Universit\'e de Toulouse, UPS, Toulouse, F-31062, France}

\author{Ludovic D. C. Jaubert}
\affiliation{CNRS, Universit\'e de Bordeaux, LOMA, UMR 5798, 33400 Talence, France}

\begin{abstract}
Skyrmions are of interest both from a fundamental and technological point of view, due to their potential to act as information carriers. But one challenge concerns their manipulation, especially at high temperature where thermal fluctuations eventually disintegrate them. Here we study the competition between skyrmions and a chiral spin liquid, using the latter as an entropic buffer to impose a quasi-vacuum of skyrmions. As a result, the temperature becomes a knob to tune the skyrmion density from a dense liquid to a diluted gas, protecting the integrity of each skyrmion from paramagnetic disintegration. With this additional knob in hand, we find at high field a topological spin glass made of zero- and one-dimensional topological defects (resp. skyrmions and bimerons).
\end{abstract}
\date{\today}
\maketitle

Experimental evidence for skyrmions is well established \cite{muhlbauer2009sk,yu2010re,munzer2010sk,woo2016,Gao2020Nat,Rosales22}. Their topological charge gives them enhanced stability and potential applications in the next-generation information storage and processing devices \cite{nagaosa2013topo,fert2013sky}. A diluted phase of skyrmion has e.g.\ been proposed as framework for a true random seed generator \cite{Pinna2018,Zazvorka19a}; a prospect that has recently spurred the search for mechanisms to stabilise such a diluted phase \cite{Mohanta2019,Kathyat2021}.

\begin{figure}[ht]
\centering\includegraphics[width=1.\columnwidth]{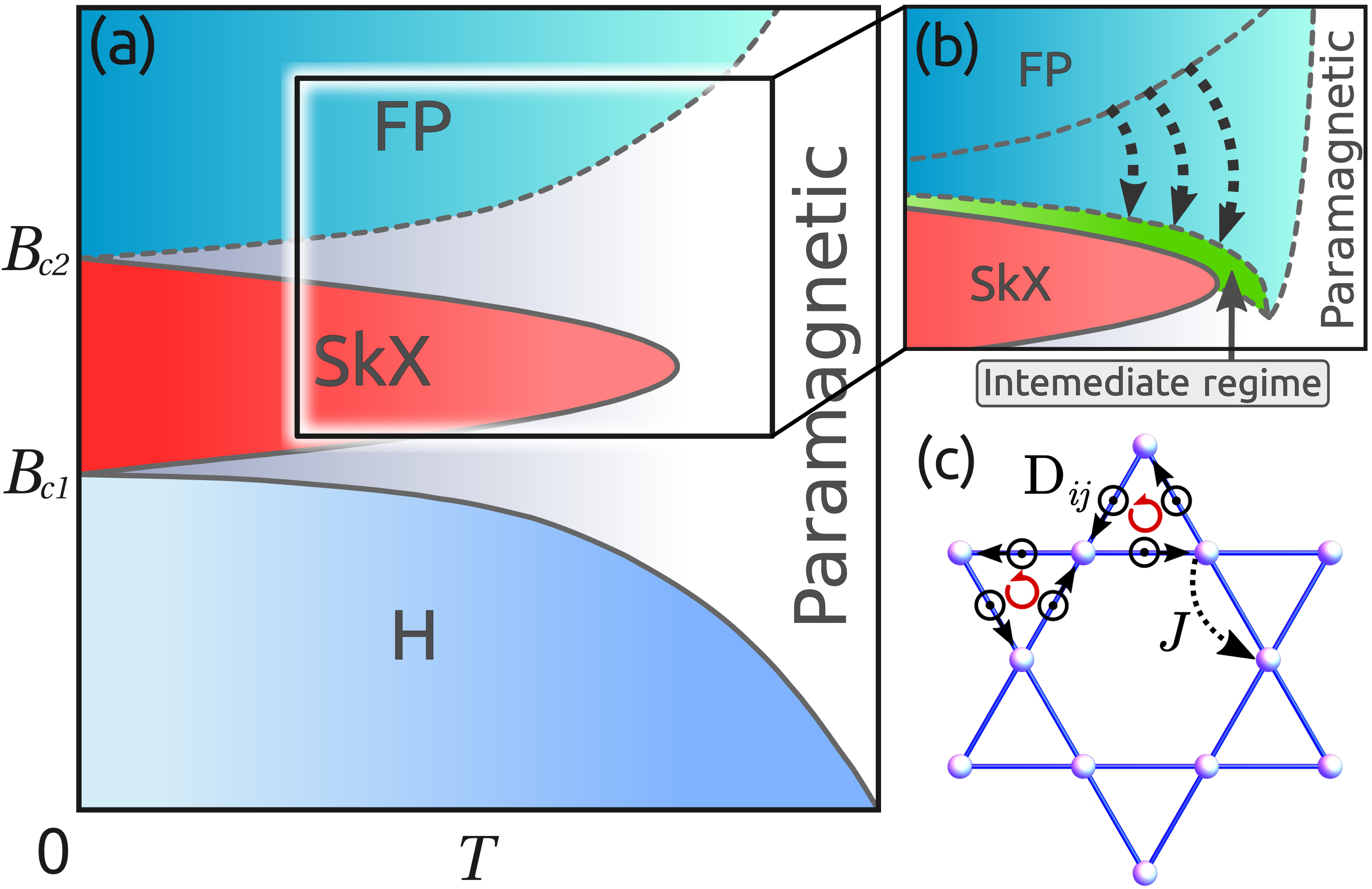}
\caption{
\textbf{(a)} Schematic representation of the traditional phase diagram where the magnetic field destroys the skyrmion solid (SkX) in favour of a field-polarised (FP) regime with no skyrmions. \textbf{(b)} Illustration of our idea where the FP phase circles around the SkX phase, thus creating an intermediate regime (in green) where skyrmion density can be tuned gradually. Solid and dashed lines respectively indicate transitions and crossovers. \textbf{(c)} Kagome lattice with nearest-neighbour Heisenberg exchange and Dzyaloshinskii-Moriya interactions.
}
\label{fig:1}
\end{figure}

The conventional route to skyrmions involves competing interactions between, for example, symmetric Heisenberg and antisymmetric Dzyaloshinskii-Moriya (DM) interactions \cite{han2010}. This competition typically favours a helical (H) phase made of one-dimensional magnetic stripes. At low but finite magnetic field $B$, these stripes break down into the famous periodic array of skyrmions (SkX), embedded in a ferromagnetic  \cite{bogdanov2001chiral,binz2006theory,roessler2006spontaneous} or antiferromagnetic \cite{rosales2015three,leonov15a,zhang2016antiferromagnetic,goebel17_antif_skyrm_cryst,osorio2019skyrmions,villalba2019field} background. At finite temperature $T$ between the H and SkX phases, this breakdown usually takes the form of elongated skyrmions called bimerons \cite{ezawa2011c}\footnote{Note that bimerons may also refer to the combination of a meron and an antimeron in in-plane magnets \cite{kharkov2017,gobel2019magnetic,jani2021af}. In this letter, bimerons always refer to ``elongated skyrmions'' and carry a net out-of-plane magnetisation opposite to the magnetic field.}. The SkX phase is then ultimately destroyed in favour of a field-polarised (FP) regime at high field.  This traditional phase diagram is illustrated in Fig.~\ref{fig:1}.(a).

The region between SkX and FP corresponds to a dilute gas of skyrmions. Since each skyrmion bears a negative magnetisation, their density can be controlled with a field $B$; a property recently used to re-visit the historical question of two-dimensional melting in a modern setting \cite{huang2020m,zazvorka2020sk,nishikawa2019s}. In contrast, heating the system at fixed field does not directly affect the number of skyrmions. Heating melts the SkX phase into a skyrmion liquid \cite{nishikawa2019s,balavz2021m,mohanta2020}, but the disappearance of skyrmions at high temperature usually comes from their individual disintegration into paramagnetic fluctuations [see video S1 of the Supplemental Material [\onlinecite{SupplementalM}]], rather than from their rarefaction within a uniformly polarised background.

Our motivation here is to propose a mechanism protecting skyrmions from paramagnetic disintegration, such that they remain well defined topological defects upon heating, all the way from the SkX melting down to quasi-zero density. To do so, we design a frustrated microscopic model where {skyrmions are separated from the paramagnetic regime by an intermediate} chiral spin liquid at finite temperature. This results in a competition between two magnetically disordered chiral phases: the spin liquid and the skyrmion fluid (the latter including two regimes, a high-density liquid and low-density gas). Finally, exploring the skyrmion fluid with two external parameters, $B$ and $T$, will see the emergence of an unconventional topological spin glass.

\noindent \textbf{Model design.} 
If we bend the FP regime downwards [Fig.~\ref{fig:1}.(b)], it will impose a quasi-vacuum of skyrmions at finite temperature, and shield the intermediate (green) region from paramagnetic fluctuations. In this green region, skyrmions would remain well defined and their density would decrease from saturation to quasi zero. But in order to bend the FP regime downwards, we need to enhance its stability at high temperature. Thermodynamically, it means increasing its entropy. We reckon it should be possible by connecting the FP regime to a classical spin liquid \cite{Knolle2019}, a phase of matter known to possess a large residual entropy. However, models for spin liquids are usually antiferromagnetic, which seems a priori incompatible with the Field-Polarised regime. Fortunately \textit{chiral} spin liquids are famous counter-examples \cite{Messio13a,he14a} where a disordered fraction of magnetic degrees of freedom co-exists with broken time-reversal symmetry.

Practically, we consider the following Hamiltonian
\begin{eqnarray}
\mathcal{H}&=&-\sum_{\langle ij\rangle}\left[J\,\Sp_i\cdot\Sp_j-\Dp_{ij}\cdot(\Sp_i\times\Sp_j)\right]-B\sum_i S^{z}_i,
\label{eq:ham}
\end{eqnarray}
where $\mathbf{S}_i$ are classical Heisenberg spins of unit length ($|\Sp_i|=1$) and $\Dp_{ij}=D^z\,\mathbf{z}+D^{\perp}\,\mathbf{e}_{ij}$ is the Dzyaloshinskii-Moriya term with out-of-plane unit vector $\mathbf{z}$ and in-plane unit vector $\mathbf{e}_{ij}$ between two neighbouring sites $\langle ij\rangle$ [Fig.~\ref{fig:1}.(c)]. The energy scale is set by $J=1$ and we study this model via classical Monte Carlo simulations {at equilibrium. As such, our approach enables} to capture the non-mean-field nature of the chiral spin liquid {and is different from dynamical studies of metastable skyrmions \cite{Wild17a,Desplat18a,Malottki19a}}. Simulation details are given in \cite{SupplementalM}.

Eq.~(\ref{eq:ham}) is not a simple bi-axial DM model. For $D^{z}=0$, it is a traditional Hamiltonian hosting magnetic skyrmions \cite{han2010,pereiro2014topological}. But for $D^{z}=\sqrt{3}$ and $D^{\perp}=B=0$, the ground state is a chiral spin liquid that combines the extensive degeneracy of the three-colouring kagome problem for in-plane spin components, with out-of-plane magnetisation \cite{Essafi2016k,Essafi2017}. Hence, in order to combine both physics, we fix $D^{z}=\sqrt{3}$ and choose $D^{\perp}=0.5$ which we find to be large enough to support skyrmion physics. The key idea is that even if skyrmion physics is stabilised at low field because of $D^{\perp}$, we expect the spin liquid to reappear at high field -- in the FP regime -- because of the Zeeman coupling to its out-of-plane magnetisation.

\begin{figure}[ht]
\centering\includegraphics[width=1\columnwidth]{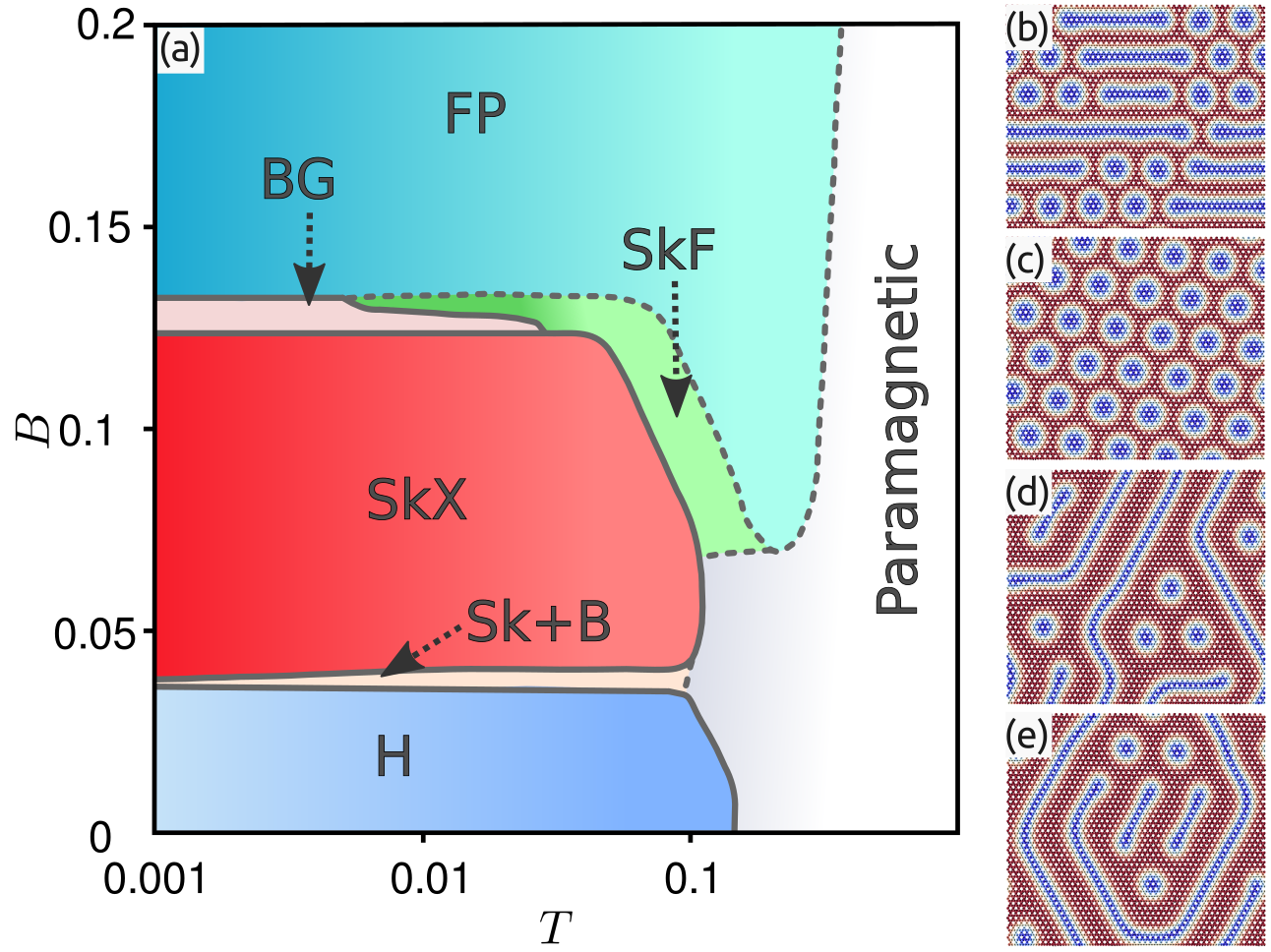}
\caption{
\textbf{(a)} Phase diagram of Hamiltonian (\ref{eq:ham}) at $D_{z}=\sqrt{3}$ and $D^{\perp}=0.5$. The field-polarised (FP) phase is a resurgence of the chiral spin liquid. Thanks to its large entropy, it circles around the SkX phase at high temperature, creating an intermediate skyrmion fluid (SkF) where the density of skyrmion can be tuned with temperature. Our model further supports a ``bimeron glass'' (BG). Solid and dashed lines respectively indicate thermodynamic transitions and crossovers, with the caveat that simulations become glassy in the vicinity of the bimeron glass.
Snapshots of spin configurations for different regions of the phase diagram: \textbf{(b)} the standard bimeron + skyrmion regime at $B=0.04$ (Sk+B) \cite{ezawa2011c}, \textbf{(c)} the skyrmion solid at $B=0.06$ and \textbf{(d,e)} the bimeron glass at $B=0.13$.
}
\label{fig:2}
\end{figure}

\begin{figure*}[ht]
\centering\includegraphics[width=1.7\columnwidth]{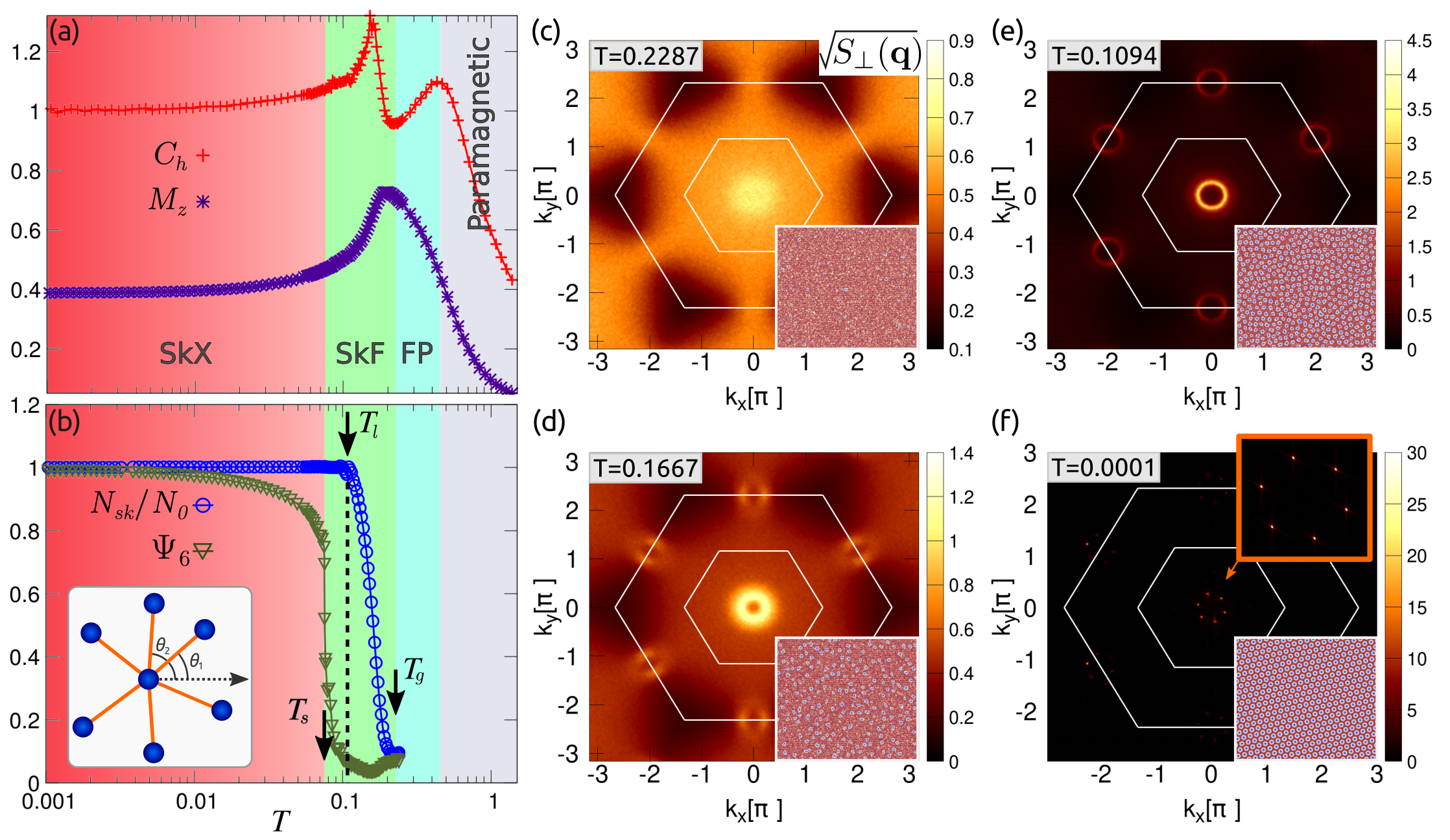}
\caption{
Competition between the chiral spin liquid and skyrmion solid at $B=0.09$ as measured from: \textbf{(a)} the specific heat $C_{h}$ and magnetisation $M_{z}$, \textbf{(b)} the normalised number of skyrmion $N_{sk}$ and orientational order parameter $\Psi_{6}$ of Eq.~(\ref{eq:psi6}) whose angle $\theta_{ij}$ is defined in the inset. $N_{0}$ is is the saturated number of skyrmions for $B=0.09$.
Equal-time structure factor $\sqrt{S_{\perp}({\bf q})}$ for different temperatures at $B=0.09$ showing the characteristic broad pinch points \textbf{(c)}, half moons \textbf{(d)}, circles \textbf{(e)} and Bragg peaks \textbf{(f)} of the field-polarised chiral spin liquid and skyrmion gas/liquid/solid respectively. Insets display snapshots of spin configurations.
}
\label{fig:3}
\end{figure*}

To quantify the non-trivial correlations of the disordered phases, we will compute the structure factor
\begin{eqnarray}
S_{\perp}({\bf q})&=&\frac{1}{N_c}\sum_{a=x,y}\langle|\sum_j S^{a}_{j}e^{i{\bf q}\cdot{\bf r_{j}}}|^2\rangle,
\end{eqnarray}
for in-plane spin components to avoid the trivial Bragg peaks coming from magnetisation. Then, with $N_{sk}$ the number of skyrmions, we define the orientational order parameter to measure the onset of SkX order \cite{nishikawa2019s,huang2020m,zazvorka2020sk,balavz2021m}
\begin{eqnarray}
\Psi_6&=&\left|\frac{1}{N_{sk}}\sum_{i=1}^{N_{sk}}\frac{1}{n_i}\sum_{j=1}^{n_i}e^{i\,6\theta_{ij}}\right|
\label{eq:psi6}
\end{eqnarray}
where $\theta_{ij}$ is defined in the inset of Fig.~\ref{fig:3}.(b) and $n_i$ is the number of nearest neighbours (labelled $j$) surrounding skyrmion $i$ as determined by Delaunay triangulation \cite{SupplementalM}. For the triangular geometry of SkX [Fig.~\ref{fig:2}.(c)], $n_{i}=6$ and $\theta_{ij}=\pi/3 $ implies that $\Psi_{6}=1$.\\

\noindent \textbf{Results.} 
One of the main outcome of our paper is that the idea presented in Fig.~\ref{fig:1}.(b) essentially works ! At low temperature, the low-field H phase gives way to the skyrmion solid (SkX) for $0.04 \lesssim B \lesssim 0.13$, as expected for a traditional skyrmion phase diagram. But at high temperature, the FP regime bends down and encircles the SkX phase as envisioned. Let us now explain the underlying mechanism in Fig.~\ref{fig:3} for $B=0.09$.

The specific heat $C_{h}$ presents two peaks [Fig.~\ref{fig:3}.(a)]. Upon cooling, the first one is accompanied by an increase of the magnetisation $M_{z}$, indicating the crossover from paramagnetism to FP regime \footnote{Since the magnetic field breaks time-reversal symmetry, the FP regime is connected to the paramagnetic phase via a crossover, and not a phase transition.}. To confirm the presence of the spin liquid in the FP regime, we notice that $C_{h}$ becomes smaller than 1 after this peak, an indication of soft modes frequently observed in spin liquids with continuous spins \cite{Chalker92a}. Furthermore, the corresponding structure factor in Fig.~\ref{fig:3}.(c) shows broad pinch points, a known signature of the flat-band physics of this spin liquid \cite{Essafi2016k}, here broadened by thermal fluctuations.

Then $M_{z}$ presents a sharp upturn at the same temperature $T_{g}$ than $C_{h}$ reaches its minimum. $T_{g}$ marks the end of the FP regime and the apparition of a gas of skyrmions, whose number $N_{sk}$ increases smoothly until reaching saturation at $T_{l}$ [Fig.~\ref{fig:3}.(b)]. Interestingly, the rise of skyrmions in real space sees the apparition of semi-circular patterns in reciprocal space [Fig.~\ref{fig:3}.(d)]. These patterns are commonly known as ``half-moons'' \cite{guitteny13a,udagawa16a} and have been observed both in neutron-scattering experiments and in theory, for a variety of frustrated systems \cite{robert08a,guitteny13a,udagawa16a,rau16b,mizoguchi2018m,yan2018half,zhang19a,Pohle21a,Gen22}. When observed in the equal-time structure factor, as it is the case in Fig.~\ref{fig:3}.(d), half moons indicate that we are in a regime dominated by low-energy dispersive bands, and at proximity from flat bands attached to a spin liquid at higher energy \cite{mizoguchi2018m}. In other words, flat bands play the role of an entropic buffer here, that stabilise the corresponding chiral spin liquid above $T_{g}$. But because they are brought to finite energy by the $D^{\perp}$ term in our model, cooling ultimately depopulates the flat bands, and the spin liquid disappears below $T_{g}$. This flat-bland depopulation is accompanied by an entropy loss marked by the second peak in specific heat. Hence, while the first peak of $C_{h}$ marks the onset of out-of-plane magnetisation, the second one marks the loss of the in-plane spin fluctuations from the spin liquid, in favour of the rise of the skyrmion population.


As $N_{sk}$ reaches saturation at $T\approx T_{l}$, the half moons ``close'' into a complete circle [Fig.~\ref{fig:3}.(e)], as expected for a dense liquid \cite{huang2020m,mohanta2020,balavz2021m}. Skyrmions do not immediately form an ordered array though. In Fig.~\ref{fig:3}.(b), $\Psi_{6}$ shows that the SkX phase only appears at $T_{s}$, offering a range of temperature, $T_{s} < T < T_{l}$, for a dense liquid of skyrmions without any long range order. Below $T_{s}$, the systems orders, as shown by the well-known Bragg peaks with six-fold symmetry of Fig.~\ref{fig:3}.(f) \cite{huang2020m,mohanta2020,balavz2021m}.\\

\noindent \textbf{Running summary.} 
Eq.~(\ref{eq:ham}) with $D^{z}=0$ is a standard skyrmion Hamiltonian whose phase diagram is schematically given in Fig.~\ref{fig:1}.(a). When $D^{z}=0$, all other parameters kept equal ($D^{\perp}=0.5, B=0.09$), video S1 in \cite{SupplementalM} shows how skyrmions are formed out of paramagnetic fluctuations {upon cooling} and already at high density once they are fully shaped; there is no diluted-gas regime.

Adding $D_{z}=\sqrt{3}$ moves Hamiltonian (\ref{eq:ham}) to the proximity of a chiral spin liquid \cite{Essafi2016k}, and modifies the phase diagram into Fig.~\ref{fig:2}.(a) where the FP regime bends down above the SkX phase. This is because once the system is magnetised out of the plane by the magnetic field, it becomes entropically favourable for in-plane spin components to fluctuate predominantly within the chiral-spin-liquid manifold [Fig.~\ref{fig:3}.(c)]. This is what confers better stability to the FP regime at high temperature, {hence separating the skyrmions from the paramagnet.} Video S2 in \cite{SupplementalM} shows the cooling process from the chiral spin liquid down to the skyrmion solid; the only difference with video S1 is that $D^{z}=\sqrt{3}$. In both videos, the skyrmion solids are the same, and to some extent, even the skyrmion liquids are very similar. However, for $T_{l} < T < T_{g}$ in our model (video S2 and Fig.~\ref{fig:3}.(b)), the number of skyrmions $N_{sk}$ gradually decreases down to almost zero because they are prevented {by the uniform magnetisation of the chiral spin liquid. As envisioned in the introduction, we were thus able to tune the density of well-defined skyrmions with temperature, and to make them vanish before being disintegrated by paramagnetic fluctuations. The spin liquid acts like a shield for the diluted-gas regime.}

Beyond skyrmion physics, we should emphasise it is noticeable for a system to visit \textit{three} different disordered magnetic regimes upon cooling: the spin liquid, skyrmion gas and skyrmion liquid, with distinct structure factors [Fig.~\ref{fig:3}.(c,d,e)]. This is why we should now take advantage of having access to two external parameters, $B$ and $T$, to explore the skyrmion fluid.\\

\noindent \textbf{Bimeron glass.} 
Bimerons are known to exist at low field, as an intermediate step during the fragmentation of the helical stripes before the SkX crystallisation \cite{ezawa2011c} [see Sk+B in Fig.~\ref{fig:2}.(a)]. They are generally not expected at high field since their magnetisation is opposite to the magnetic field. This is, however, what we observe at $B\approx 0.13$ [see BG in Fig.~\ref{fig:2}.(a)]. As the SkX melts, skyrmions become unstable and split into two merons, forming extended bimerons in the shape of ``snakes'' \cite{zhang2018,derras2021,chakrabartty2021} [Fig.~\ref{fig:2}.(d,e)]. Our understanding is that the number of skyrmions is too low to form a solid, and the system dynamically adapts by filling the gap with extended bimerons. Colloquially speaking, it is as if the bimerons were ``connecting the dots'' of the triangular SkX array.

Once nucleated, the growth of bimerons is almost instantaneous, suggesting out-of-equilibrium metastable effects [see video S3 in \cite{SupplementalM}]. Indeed, using Landau-Lifshitz dynamics \cite{gerling90a,robert08a,Bilitewski19a}, the autocorrelation
\begin{eqnarray}
A(t)=\dfrac{\frac{1}{N}\sum_{i}\langle \mathbf{S}_{i}(t)\cdot\mathbf{S}_{i}(0)\rangle-M(t)\;M(0)}{1-M(t)\;M(0)}
\label{eq:At}
\end{eqnarray}
of this regime becomes $\sim 6$ orders of magnitude slower than in the skyrmion fluid. This unconventional glassiness, for a system \textit{without} quenched disorder, is remarkable, and a direct consequence of the macroscopic size and topological stability of the bimerons. We coin this regime a bimeron glass (BG), made of zero- and one-dimensional topological defects. This frozen arrangement of bimerons and skyrmions is similar to Hamiltonian chains \cite{jacobsen2007,footnote}, with an extensive, countable, degeneracy and potential to store information. In that context, the frozen dynamics offers a welcome topological stability to the spin configurations.\\

\begin{figure}[h]
\centering\includegraphics[width=0.9\columnwidth]{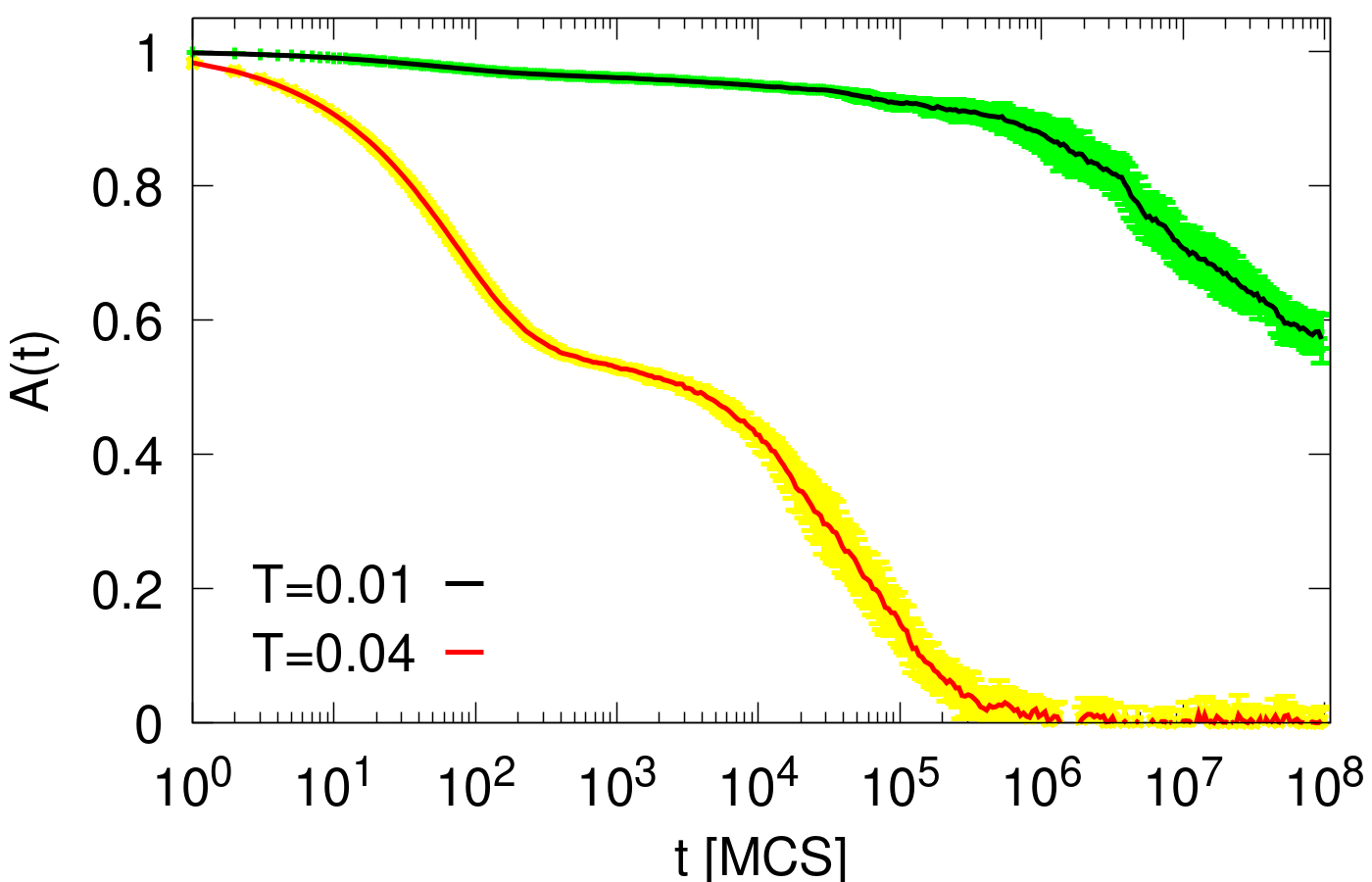}
\caption{
Autocorrelation $A(t)$ as a function of time $t$ from Landau-Lifshitz dynamics at $B=0.13$ in the skyrmion fluid ($T=0.04$ in red) and bimeron glass ($T=0.01$ in black). The latter is $\sim 6$ orders of magnitude slower than the former. Error bars are coming from averaging over different seed configurations obtained from Monte Carlo simulations.
}
\label{fig:4}
\end{figure}

\noindent \textbf{Discussion.}
Our work can be seen as a proof of concept. Since the field-polarised phase is always present at high field, we expect our results to be valid beyond traditional skyrmion models, in two or three dimensions, as long as a chiral spin liquid can be found at proximity in parameter space. In that sense, exploring the competition between complex skyrmion models (e.g.\ \cite{leonov15a,rozsa2016complex,bottcher2018b}) and other spin liquids with extensive residual entropy (i.e.\ different from spiral spin liquids \cite{gao17}) would be a useful endeavour for potential experimental realisations. Synthesising a material described by fine tuned parameters is always a challenge, but the advantage of a spin liquid is that its entropic influence is usually felt at finite temperature over a broad region in parameter space, away from the pristine Hamiltonian.

The bimeron glass also deserves further investigation in itself: e.g.\ to explore its metastability, to see if bimeron configurations can be manipulated, or to understand it in a broader picture (e.g.\ with polymer thin films \cite{abate2021}). Since our model supports several chiral magnetic textures, one can also expect unconventional behaviours when coupled to itinerant electrons, such as anomalous Hall effects. And since the density of skyrmions varies rapidly with temperature, would it be possible to induce a skyrmion current in a temperature gradient?

To conclude, let us mention that the chiral spin liquid studied here can be mapped exactly onto a specific point XXZ$_{0}$ of the XXZ Hamiltonian on the kagome lattice \cite{Essafi2016k}. Since this mapping only transforms in-plane spin components, it means that the magnetic field along the $z-$axis is invariant. Hence, all the physics presented in this paper can be exactly transposed onto a XXZ kagome model with rotated in-plane Dzyaloshinskii-Moriya interactions and $D^{z}=0$. The consequences of the remarkable properties of the quantum XXZ$_{0}$ model \cite{changlani2018,Lee2020} on the skyrmion fluid is a vast and open question.\\

\noindent \textbf{Acknowledgments -} 
We thank L. Cugliandolo, S. Grigera and P. Holdsworth for useful comments.
H.D.R. and F.A.G.A. acknowledge financial support from CONICET (PIP 2021-11220200101480CO) and SECyT-UNLP (I+D X893), and F.A.G.A. from PICT 2018-02968, and L.D.C.J. from ANR-18-CE30-0011-01.

\bibliographystyle{apsrev4-1}
%

\begin{thebibliography}{75}
\bibitem [{foo()}]{footnote}%
  \BibitemOpen
  \href@noop {} {\enquote {\bibinfo {title} {Note that we have extended
  {J}acobsen's definition of hamiltonian chains to include single vertices,
  i.e. skyrmions.}}\ }\BibitemShut {NoStop}%
\end{thebibliography}
\end{document}